\documentclass{article}
\usepackage{spconf}
\usepackage[utf8]{inputenc}
\usepackage{amsfonts}
\usepackage{amssymb}
\usepackage{amsmath}
\usepackage{microtype}
\usepackage{graphicx}
\usepackage{booktabs}
\usepackage{hyperref}
\usepackage{multirow}
\usepackage{footnote}
\usepackage{tablefootnote}
\usepackage{siunitx} 
\usepackage{xcolor} 
\usepackage{gensymb} 
\usepackage{commath} 
\usepackage{bm} 

\makesavenoteenv{table}
\makesavenoteenv{table*}
\makesavenoteenv{tabular}

\newcommand{\R}[1] {\textnormal{#1}}

\title{Perception through 2D-MIMO FMCW Automotive Radar Under Adverse Weather}

\name{Xiangyu Gao, Sumit Roy, Guanbin Xing, Sian Jin}
\address{\{xygao, sroy, gxing, sianjin\} @uw.edu \\ Department of Electrical and Computer Engineering, University of Washington}

\begin{document}

\setlength{\abovedisplayskip}{3pt}
\setlength{\belowdisplayskip}{3pt}

\maketitle

\begin{abstract}
Millimeter-wave (mmWave) radars are being increasingly integrated in commercial vehicles to support new Adaptive Driver Assisted Systems (ADAS) features that require accurate location and Doppler velocity estimates of objects, independent of environmental conditions. To explore radar-based ADAS applications, we have updated our test-bed with Texas Instrument's mmWave cascaded FMCW radar (TIDEP-01012) that forms a non-uniform 2D MIMO virtual array. In this paper, we develop the necessary received signal models for applying different direction of arrival (DoA) estimation algorithms and experimentally validating their performance on formed virtual array under controlled scenarios. To test the robustness of mmWave radars under adverse weather conditions, we collected raw radar dataset (I-Q samples post demodulated) for various objects by a driven vehicle-mounted platform, specifically for snowy and foggy situations where cameras are largely ineffective. Initial results from radar imaging algorithms to this dataset are presented. 
\end{abstract}


\section{Introduction}
\label{sec:intro}

To meet requirements for ADAS and especially L4/L5 autonomous driving \cite{6Levels}, automotive radars need to have a high angular resolution. There are several ways of improving radar angular resolution: 1) using more physical antenna elements, or a non-uniform array with larger antenna distance; 2) increasing the antenna aperture via synthetic aperture radar \cite{gao2021mimosar} concepts that exploit vehicle-mounted radar movement; 3) forming virtual array via multiple-input and multiple-out (MIMO) radar operations \cite{ti_mimo, 9266601}. In MIMO radar, multiple transmit (TX) antennas send orthogonal signals, which enables the contribution of each TX signal to be extracted at each receive (RX) antenna. Hence a physical TX array with $M_\R{T}$ elements and RX array with $M_\R{R}$ elements will result in a virtual array with upto $M_\R{T}M_\R{R}$ unique (non-overlapped) virtual elements \cite{Wang2012VirtualAA}. To reduce array cost (fewer physical antenna elements), non-uniform arrays spanning large apertures, e.g., minimum redundancy array (MRA) \cite{mra} have been proposed. 

\begin{figure}
\vspace{-2em}
\centering
\includegraphics[width=0.5\textwidth, trim=1 2 1 1,clip]{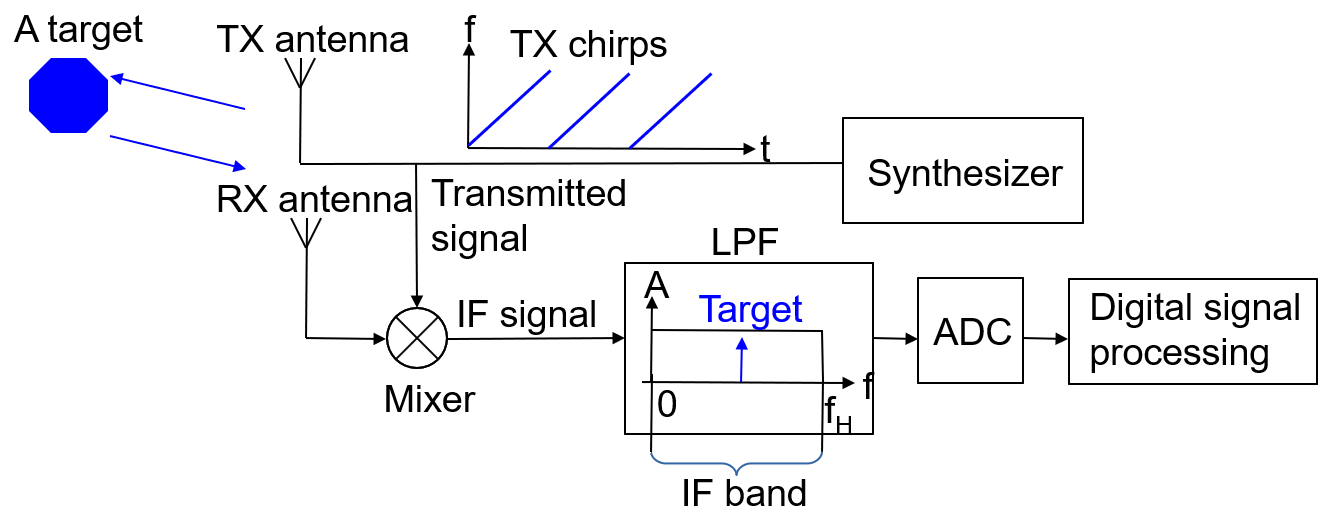}
\vspace{-2em}
\caption{Basic FMCW radar block diagram}
\label{fig:fmcw}
\vspace{-1.5em}
\end{figure}

From the received signals at RX array elements, the DoA of targets can be extracted by proper signal processing. The FFT-based DoA method and the multiple signal classification (MUSIC) are discussed and experimented in \cite{exper} on a 2 TX and 4 RX MIMO radar. Compressive sensing (CS) methods have been exploited in DoA estimation to exploit the inherent spatial sparsity of targets, via recovery algorithms from spatially under-sampled measurements \cite{YANG2018509}. For our application, CS-based DoA is applied to \textit{non-uniformly spaced array}, that potentially enable higher resolution by reconstructing the observations at the missing elements \cite{8698136}. Further, CS is known to mitigate the high sidelobes originating from non-uniform array \cite{1137832} and therefore reduce false alarms \cite{8448197, 8714419}.

While mmWave radars are generally known for excellent environmental robustness under adverse weather conditions \cite{Zang2019TheIO}, there has been little published studies to date experimentally verifying this hypothesis due to the difficulty of capturing such data. Prior work \cite{fogexp} studied the effect of fog on the mmWave propagation, and Gao et al. \cite{ramp} showed a robust and high-performance object recognition algorithm verified on nighttime data where cameras are largely ineffective. In this paper, we present a new CS-based DoA algorithm, whose performance is validated for data obtained using a frequency-modulated continuous wave (FMCW) \SI{77}{GHz} radar test platform that enables a non-uniform 2D MIMO virtual array. In addition, we present some initial results regarding operational robustness to inclement weather by stress-testing performance of DoA on a collected dataset for snowy and foggy conditions.

\vspace{-0.5em}
\section{FMCW MIMO Radar}
\vspace{-0.5em}
\subsection{FMCW Radar and Range Estimation}
FMCW radar transmits periodic wideband linear frequency-modulated (LFM, also called chirps) signal as shown in Fig.~\ref{fig:fmcw}. The TX signal is reflected from targets and received at the radar receiver. FMCW radars can detect targets' range and velocity from the RX signal using the stretch or de-chirping processing structure \cite{exper} in Fig.~\ref{fig:fmcw}. A mixer at the receiver multiplies the RX signal with the TX signal to produce an intermediate frequency (IF) signal. Since the RX and the TX signal are both LFM signal with constant frequency difference determined by target's location, the IF signal is a single-tone signal.
For example, the IF signal for a target at range $r$ has frequency $f_\R{IF} = \frac{2r}{c}S$, the multiplication of round-trip delay $\frac{2r}{c}$ with chirp slope $S$, where $c$ denotes the speed of the light. Thus, detecting the frequency of the IF signal can solve the target range. At the end of receiver, IF signal is passed into an anti-aliasing low-pass filter (LPF) and an analog-to-digital converter (ADC) for following digital signal processing. A fast Fourier transform (FFT) is widely adopted to estimate $f_{\R{IF}}$ to infer $r$, and hence such operation is called the \textit{Range FFT}.

\begin{figure}[t]
\centering
\includegraphics[width=0.35\textwidth, trim=1 2 1 1,clip]{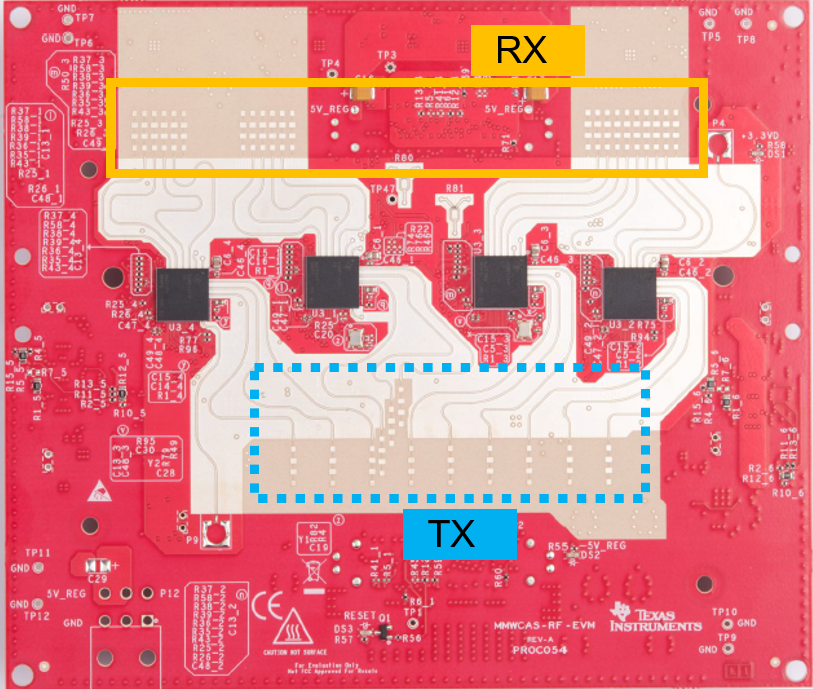}
\caption{Texas Instrument 4-chip cascaded radar board \cite{ti_casd} and the position of antennas.}
\label{fig:cascd}
\end{figure}

\begin{figure}[t]
\centering
\includegraphics[width=0.48\textwidth, trim=1 2 1 1,clip]{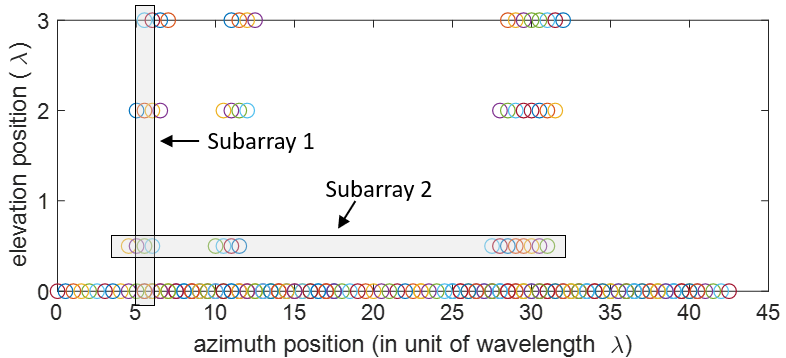}
\caption{2D MIMO virtual array formed by 12 TX and 16 RX.}
\label{fig:virt}
\vspace{-1em}
\end{figure}

\subsection{MIMO Radar and Virtual Array}

To estimate the direction of angle of targets relative to a receiver orientation, an antenna array is needed. In MIMO radar, a virtual array located at the spatial convolution of TX antennas and RX antennas is enabled by the orthogonality of TX signal \cite{mimo_advanced}. The convolution produces a set of virtual element locations that is the sum of the TX and RX element locations. For example, if an automotive radar consists of a RX linear array of $M_{\R{R}}$ elements with $\lambda / 2$ spacing combined with a TX array of 2 elements which are spaced $M_{\R{R}} \lambda / 2$ apart, the synthesized MIMO virtual array is a $2 M_{\R{R}}$-elements uniform linear array (ULA) with $\lambda / 2$ spacing.


We adopt TIDEP-01012 \cite{ti_casd}, a high-resolution mmWave FMCW radar board composed of four AWR2243 chips from Texas Instrument (TI) for experiments. This radar includes 12 TX and 16 RX antennas placed in specific 2D manner shown in Fig.~\ref{fig:cascd}, which creates a 2D virtual array (Fig.~\ref{fig:virt}) with 192 elements via the spatial convolution of all TX and RX. The resulting virtual array has some overlapped elements and is mostly sparse except the bottom row (a ULA with 86 elements). For processing, we selected data for 2 subarrays - the vertical subarray 1 is a MRA \cite{mra} with 4 non-uniform spacing elements spanning $3\lambda$ aperture, and the non-uniform horizontal subarray 2 with 16 elements spanning $26.5\lambda$.

\begin{figure}[!h]
\centering
\includegraphics[width=0.4\textwidth, trim=1 2 1 1,clip]{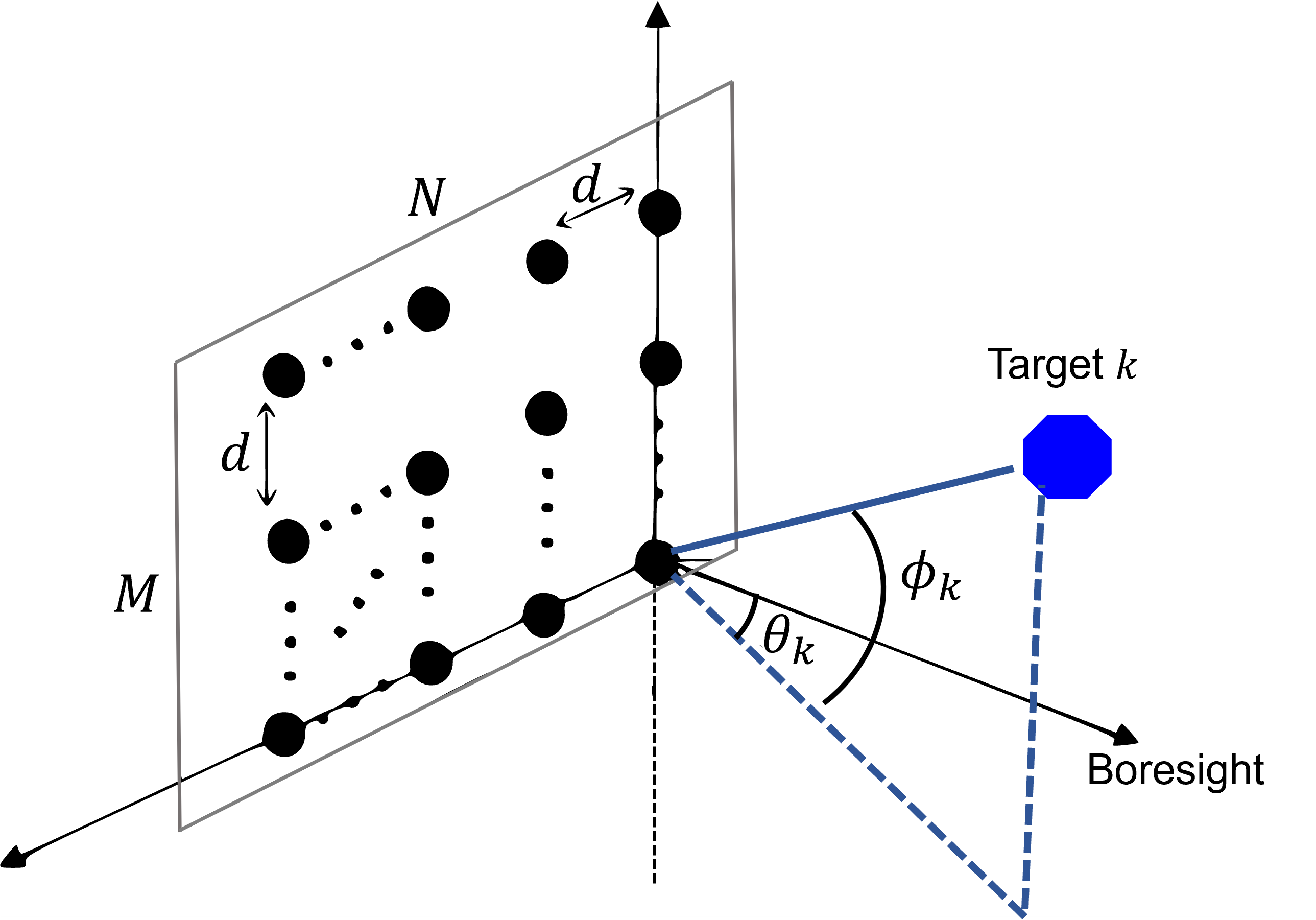}
\caption{System model of the uniform plane array.}
\label{fig:sys_model}
\vspace{-1em}
\end{figure}

\begin{table*}[!ht]
\begin{center}
\caption{Parameter calculation (based on \cite{exper}) and configuration for 4-chip cascaded radar test-bed}
\resizebox{0.95\textwidth}{!}{
\begin{tabular}{ll|ll}
\toprule  
Parameter & Calculation Equation & Configuration & Value\\
\midrule  
Range resolution ($R_{\R{res}}$) & $R_{\R{res}}=\frac{c}{2B}=\SI{0.39}{m}$  &  Frequency ($f_{\R{c}}$) & \SI{77}{GHz}\\[0.75ex]

Velocity resolution ($V_{\R{res}}$) &  $V_{\R{res}}=\frac{\lambda}{2N_{\R{c}} T_{\R{c}}}=\SI{0.0631}{m/s}$ & Sweep Bandwidth ($B$) & \SI{384}{MHz}\\[0.75ex]

Azimuth Angle resolution ($\theta_{\R{res}}$) ~\protect \footnotemark  &  $\theta_{\R{res}}=\frac{\lambda}{L_\R{h}\cos{\theta}} \approx \ang{1.35}$  & Sweep slope ($S$) & \SI{45}{MHz \per\micro\second} \\[0.75ex]

Elevation Angle resolution ($\phi_{\R{res}}$)~\protect \footnotemark  &  $\phi_{\R{res}}=\frac{\lambda}{L_\R{v}\cos{\theta}} \approx \ang{19}$  & Sampling frequency ($f_{\R{s}}$) & \SI{15}{Msps}\\[0.75ex]

Max operating range ($R_{\R{max}}$) & $R_{\R{max}} = \frac{f_{\R{s}} c}{2S} = \SI{50}{m}$ & Num of chirps in one frame ($N_{\R{c}}$) & $128$ \\[0.75ex]

Max operating velocity ($V_{\R{max}}$) & $V_{\R{max}} = \frac{\lambda}{4T_{\R{c}}} = \SI{4.04}{m/s}$ & Num of samples of one chirp ($N_{\R{s}}$) & $128$ \\[0.75ex]

& & Duration of chirp~\protect \footnotemark~and frame ($T_{\R{c}}$, $T_{\R{f}}$) & \SI{240}{\micro\second}, \SI{1/30}{s} \\
\bottomrule 
\label{tab:sys_param}
\end{tabular}}
\end{center}
\vspace{-2em}
\end{table*}

\footnotetext[1]{$\theta_{\R{res}}$ is determined by maximum horizontal aperture length $L_\R{h}=42.5\lambda$.}

\footnotetext[2]{$\phi_{\R{res}}$ is determined by the maximum vertical aperture length $L_\R{v}=3\lambda$.}

\footnotetext{$T_\R{c}$ is equal to chirp interval times number of TX antennas.}

\section{System Model for DoA Estimation}

Without loss of generality, we consider a RX uniform plane array (UPA) in the vertical plane with $M (N)$ antenna elements in each row (column), respectively. The array response is given by \cite{8587926}:
\begin{align}
\mathbf{y}=\mathbf{A} \mathbf{x}+\mathbf{n}
\end{align}

\noindent where $\mathbf{n}$ is a noise term, $\mathbf{x}=\left[\beta_{1}, \ldots, \beta_{K}\right]^{\R{T}}$ is the reflection coefficient matrix for $K$ targets, and $\mathbf{A}=\left[\mathbf{a}_1, \ldots, \mathbf{a}_K\right]$ is the array steering
matrix with
$$
\mathbf{a}_k=\mathbf{a}\left(u_k\right) \otimes \mathbf{a}\left(v_k\right)
$$
$$
\mathbf{a}\left(u_k\right)=\left[1, e^{j \frac{2 \pi}{\lambda} d \sin \phi_{k}}, \ldots, e^{j\frac{2 \pi}{\lambda} \left(N-1\right) d \sin \phi_{k}}\right]^{\R{T}}
$$
$$
\mathbf{a}\left(v_k\right)=\left[1, e^{j\frac{2 \pi}{\lambda} d \sin \theta_{k} \cos \phi_{k}}, \ldots, e^{j\frac{2 \pi}{\lambda} \left(M-1\right) d \sin \theta_{k} \cos \phi_{k}}\right]^{\R{T}}
$$

Here, $\mathbf{a}\left(u_k\right)$ and $\mathbf{a}\left(v_k\right)$ are steering vectors for elevation angle $\phi_k$ and azimuth angle $\theta_k$ for $k$th target, respectively. $\otimes$ denotes the Kronecker product operation, and $d$ is the antenna spacing. 
For a 1D ULA, angle finding can be done with digital beamforming by performing FFT across the received signal of array elements \cite{exper}. This FFT-based method can be extended to above 2D UPA, i.e., perform first FFT on the horizontal elements and second FFT on the elevation elements, which is \textit{computationally efficient} but has low resolution.

\subsection{MUSIC}
MUSIC belongs to the class of {\em eigen-decomposition} based DoA estimators that construct the $(MN-K)$-dimension noise subspace $U_{\R{n}}$ and the left $K$-dimension signal subspace from the covariance matrix of received signals $\mathbf{y}$ \cite{MUSIC}. The azimuth and elevation angles $(\theta_k, \phi_k)$ of the $k$th target can be found as peak on 2D MUSIC spectrum, which is given by \cite{MUSIC}: 
\begin{align}
    P_{\mathrm{MUSIC}}(\theta_k,\phi_k)=\frac{1}{\mathbf{a}_k^{\R{H}} \boldsymbol{U}_{\R{n}} \boldsymbol{U}_{\R{n}}^{\R{H}} \mathbf{a}_k}
\end{align}

\subsection{Compressive Sensing (CS)}

To apply CS to DoA estimation, we need to define a search grid of $K_\R{g}\left(K_\R{g} \gg K\right)$ potential incident angles, and construct an hypothetical array steering matrix $\Tilde{\mathbf{A}}=\left[\mathbf{a}_1, \ldots, \mathbf{a}_{K_\R{g}}\right]$ and the reflection coefficient matrix $\Tilde{\mathbf{x}}=\left[\beta_{1}, \ldots, \beta_{K_\R{g}}\right]^{\R{T}}$. 

The CS-based DoA estimation problem can be solved by an $\ell_1$-norm regularized convex optimization, named square-root LASSO \cite{YANG2018509}:
\begin{equation}
\min_{\Tilde{\mathbf{x}}} \xi \|\Tilde{\mathbf{x}}\|_{1} + \|\Tilde{\mathbf{A}} \Tilde{\mathbf{x}}-\mathbf{y}\|_{2}
\end{equation}

\noindent where $\|\cdot\|_{1}$ is the $\ell_{1}$-norm forces the sparsity constraint, and $\xi > 0$ is a regularization parameter. 

Above MUSIC and CS estimator are modeled for 2D UPA, and thus can address azimuth and elevation DoA estimation together. For ease of performing experiments and testing performance, we only use them for 1D DoA estimation next.


\section{Experiments}
\subsection{Radar Test-bed and Configuration}

\begin{figure}[t]
\centering
\includegraphics[width=0.45\textwidth, trim=1 2 1 1,clip]{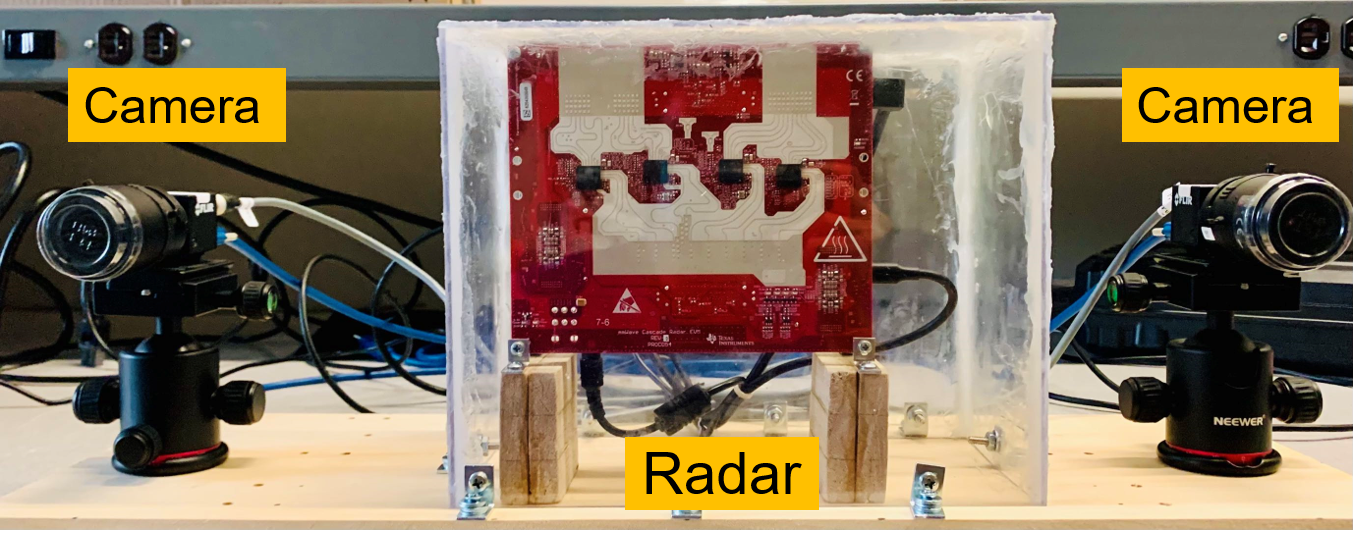}
\caption{4-chip cascaded radar test-bed with 2 cameras.}
\label{fig:platform}
\end{figure}

We assembled a test-bed (see Fig.~\ref{fig:platform}) with the TIDEP-01012 radar \cite{ti_casd} and binocular FLIR cameras (left and right). Binocular cameras are synchronized with radar to provide the visualization for the imaging scenarios. The 4-chip cascaded radar forms a large 2D-MIMO virtual array (see Fig.~\ref{fig:virt}) via the time-division multiplexing (TDM) \cite{ti_mimo} on 12 TX antennas, resulting in substantial raw data ($\sim$ \SI{378}{MB}) per second. Other configuration values of this radar are shown in Table.~\ref{tab:sys_param}.

\begin{figure}[t]
\centering
\includegraphics[width=0.45\textwidth, trim=1 2 1 1,clip]{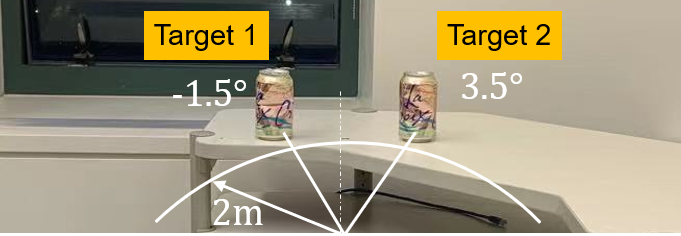}
\caption{Experiment setup with two targets separated by \SI{5}{\degree}.}
\label{fig:ref_cors}
\vspace{-1em}
\end{figure}

\begin{figure}[t]
\centering
\includegraphics[width=0.42\textwidth, trim=1 0 1 1,clip]{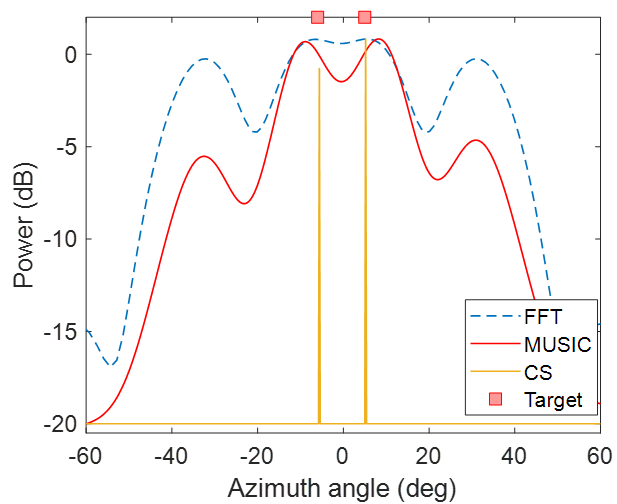}
\caption{DoA spectrums for FFT, MUSIC, and CS estimators with simulation on vertical non-uniform subarray 1 (in Fig.~\ref{fig:virt}).}
\label{fig:doa_simu}
\end{figure}

\begin{figure}[t]
\centering
\includegraphics[width=0.45\textwidth, trim=1 0 1 1,clip]{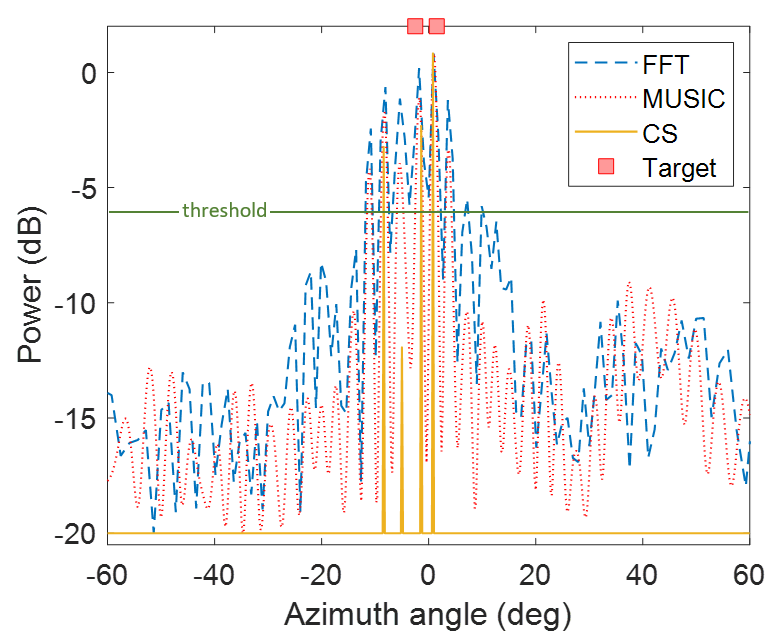}
\caption{DoA spectrums for FFT, MUSIC, and CS estimators with experiment on horizontal subarray 2 (in Fig.~\ref{fig:virt}).}
\label{fig:doa_ref_cors}
\end{figure}

\subsection{DoA Estimation on Non-uniform Array}

Since the virtual array in Fig.~\ref{fig:virt} is mostly non-uniform, we evaluate the performance of different DoA estimators with it. First, we simulate the radar received signal of vertical subarray 1 (a MRA in Fig. \ref{fig:virt}) for two point targets located at range \SI{20}{m}, \SI{-6}{\degree} and \SI{5}{\degree} elevation respectively. Three DoA algorithms - FFT, MUSIC, and CS - are implemented on the simulated signal to obtain the spectrum maps shown in Fig.~\ref{fig:doa_simu}. Note that we choose regularization parameter $\xi=1.4$ in CS reconstruction, based on exhaustive search.
The results show that MUSIC and CS-based DoA estimators achieve \textit{improved resolution} for non-uniform array by the ability to separate two targets, while FFT does not. Besides, CS generates the sparse solution that \textit{avoids high sidelobes} at around \SI{\pm 30}{\degree}. To compare the DoA estimation accuracy, we calculate the root-mean-square errors (RMSE) for all methods by averaging over 30 simulation rounds. We got the RMSE of MUSIC method (\SI{2.4609}{\degree}) and CS method (\SI{0.3162}{\degree}), which demonstrates that CS-based DoA estimation is more accurate than MUSIC on selected non-uniform linear array.

Second, we employed a setup with two close targets placed at \SI{2}{m}, \SI{-1.5}{\degree} and \SI{3.5}{\degree} azimuth respectively (see Fig.~\ref{fig:ref_cors}), and collected the real radar return signal. The power spectrum for the received signal for horizontal subarray 2 (in Fig.~\ref{fig:virt}) using our CS-DOA approach is shown in Fig.~\ref{fig:doa_ref_cors} and shows 3 strong peaks - two of them corresponds to targets, while the third is likely a spurious reflection from an indoor wall. To evaluate the false alarms of FFT and MUSIC based DoA estimation caused by non-uniform array spacing, we set a \SI{-6}{dB} threshold and count the additional peaks exceeding magnitude threshold. Results show that CS, FFT and MUSIC estimator have 0, 5 and 3 false alarms respectively, which verifies that CS is more robust for DoA estimation. It is to be noted that the performance of CS is dependant on the choice of regularization value $\xi$. The optimal regularization $\xi$ is a function of the number of targets, and may be determined by exhaustive search to find the optimal value.


\subsection{Radar Dataset and Imaging for Adverse Weathers}

\begin{figure}
\vspace{-1em}
\centering
\includegraphics[width=0.43\textwidth, trim=1 2 1 1,clip]{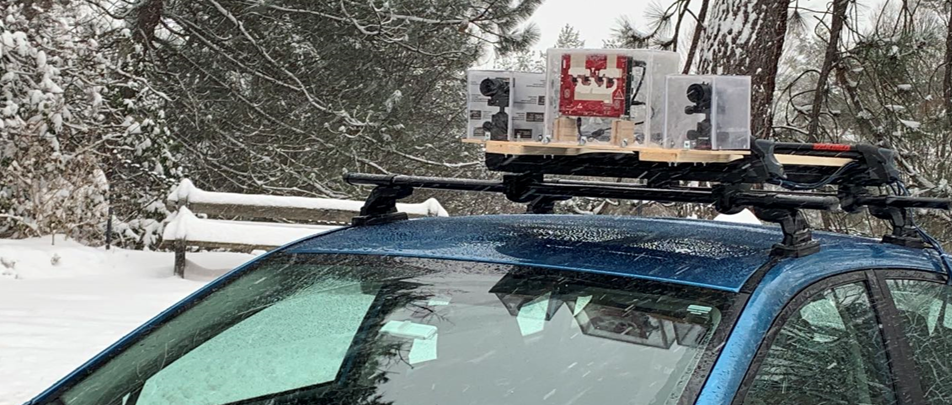}
\caption{Test-bed mounted on a vehicle for dataset collection.}
\label{fig:mount_plat}
\end{figure}

\begin{figure}
\centering
\includegraphics[width=0.50\textwidth, trim=2 2 1 1,clip]{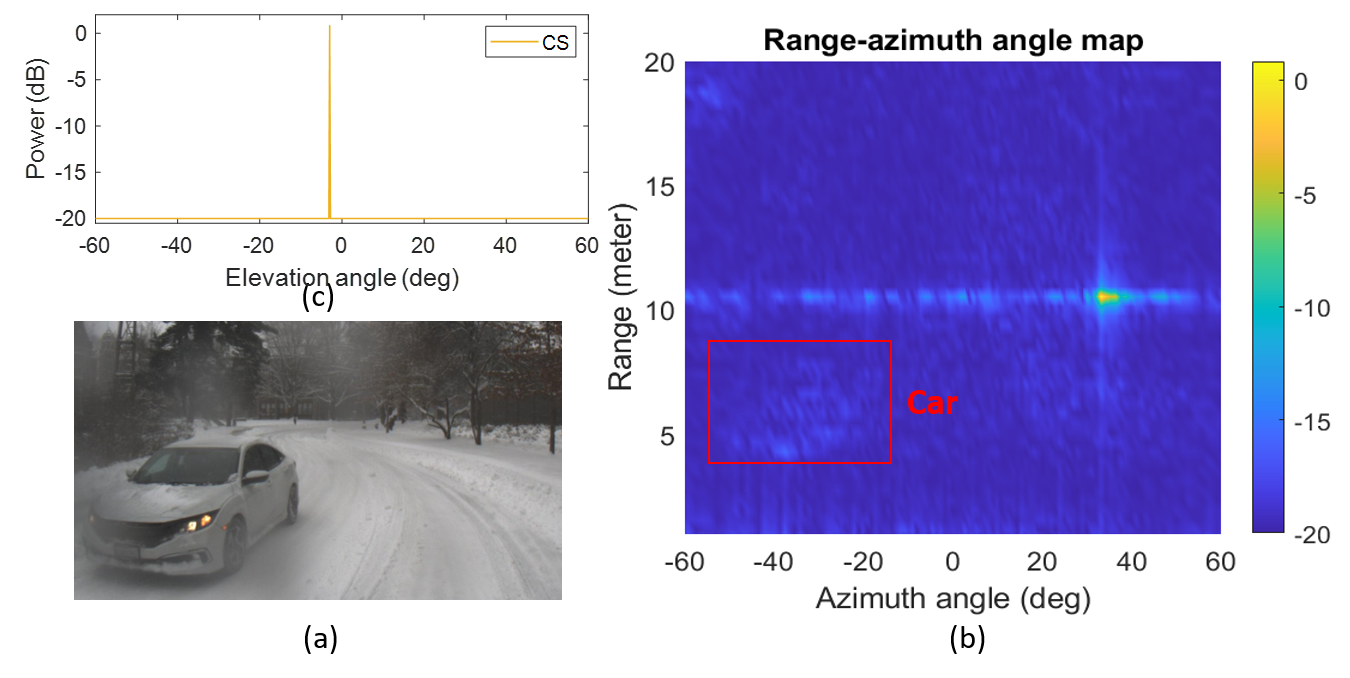}
\caption{An example of adverse-weather radar dataset: a moving vehicle with its (a) camera image, (b) range-azimuth angle map, (c) elevation DoA spectrum for range $r=\SI{4.3}{m}$.}
\label{fig:doa_roadcar}
\vspace{-1em}
\end{figure}

For promoting the development of high-level radar ADAS applications (e.g., object recognition \cite{ramp}) under critical adverse weathers, we mount the test-bed on a vehicle (see Fig.~\ref{fig:mount_plat}) and collect raw radar I-Q samples dataset for various objects (pedestrian, cyclist, and car) by driving vehicle in \textit{snowy and foggy conditions} where camera images are compromised. 

We present an example from the collected dataset in Fig.~\ref{fig:doa_roadcar}, with a camera image of a moving car and corresponding radar imaging results. The range-azimuth angle map (see Fig.~\ref{fig:doa_roadcar}(b)) is generated by performing Range FFT and FFT-based DoA on the radar data of bottom-row ULA (in Fig.~\ref{fig:virt}). We also show the elevation DoA spectrum for range \SI{4.3}{m} (in Fig.~\ref{fig:doa_roadcar}(c) with ground truth around \SI{-3}{\degree}), which is obtained by executing CS on corresponding radar data of vertical subarray 1 in Fig.~\ref{fig:virt}. According to qualitative results, the vehicle object is still visible in radar image even with the attenuation from snow and fog, which validates the robustness of mmWave radar primitively.

\section{Conclusion}
A high-resolution mmWave FMCW radar test-bed with non-uniformly spaced 2D-MIMO virtual array was used for testing a CS-DoA algorithm performance, initially calibrated and benchmarked for some test cases followed by evaluation for a dataset collected under adverse weather (snow, fog) conditions. 

\bibliographystyle{IEEEbib}
\bibliography{bibtex}

\end{document}